\documentclass[aip,graphicx]{revtex4-2}
\draft 

 \usepackage{graphicx}  
  \usepackage{xcolor}  \usepackage[colorlinks,citecolor=blue,linkcolor=blue,urlcolor=blue,bookmarks=false,hypertexnames=true]{hyperref}

\usepackage{graphicx}
\usepackage{stackengine}
\usepackage{dcolumn}
\usepackage{bm}
\usepackage{xcolor}
\usepackage{mathtools}
\usepackage{bbold}
\usepackage{subfigure}
\usepackage{mathtools} 
\usepackage{nicefrac}
\usepackage{pgfplots}
\usepackage{float}
\usepackage{comment}
\usepackage{makecell}

\usepgfplotslibrary{external}
\pgfplotsset{compat=1.18}
\newcommand{\ket}[1]{| #1 \rangle}
\newcommand{\bra}[1]{\langle #1 |}

\newcommand{\ketbra}[2]{\left|#1\right\rangle\hspace{-1.1mm}\left\langle #2 \right|}

\begin{document}


\title{Suppression of coherent errors during entangling operations in NV centers in diamond} 

\author{Regina Finsterhoelzl}
\email{regina.finsterhoelzl@uni-konstanz.de}
\author{Guido Burkard}
\email{guido.burkard@uni-konstanz.de}
\affiliation{Department of Physics, University of Konstanz, 78462 Konstanz, Germany}

\date{\today}

\begin{abstract}
We consider entangling operations in a single nitrogen-vacancy (NV) center in diamond where the hyperfine-coupled nuclear spin qubits are addressed with radio-frequency (rf) pulses conditioned on the state of the central electron spin. Limiting factors for the gate fidelity are coherent errors due to off-resonant driving of neighboring transitions in the dense, hyperfine-split energy spectrum of the defect and non-negligible perpendicular hyperfine tensor components that narrow the choice of $^{13}\rm C$ nuclear spin qubits. We address these issues by presenting protocols based on synchronization effects that allow for a complete suppression of both error sources in state-of-the-art CNOT gate schemes. This is possible by a suitable choice of parameter sets that incorporate the error into the scheme instead of avoiding it. These results contribute to the recent progress toward scalable quantum computation with defects in solids.
\end{abstract}

\pacs{}

\maketitle 

\section{Introduction}
Single defects in solids are intensively studied as versatile platforms for quantum science and technologies \cite{Awschalom2018, Zwanenburg2013, Weber2010, Bourassa2020, Castelletto2020, Castelletto2024, Ecker2024, Pasini2023, Karapatzakis2024}. The nitrogen-vacancy (NV) center is a promising system not only for quantum sensing \cite{Degen2017} but has recently also been employed for the realization of quantum networks \cite{Nemoto2014, Ruf2021, Humphreys2018, Pompili2021} and quantum computation \cite{Wrachtrup2006, Jelezko2006, Doherty2013, Dobrovitski2013, Pezzagna2021, Sun2023, Sekiguchi2023, Waldherr2014, Taminiau2014, Cramer2016, Finsterhoelzl2022, Ruh2022, Abobeih2022, Debone2024}.
As a point defect consisting of a single nitrogen atom next to an empty site in the diamond lattice, the long-lived electron spin of the NV center is well protected in the lattice of the diamond host material \cite{Taminiau2014, Abobeih2018} while allowing optical initialization and readout \cite{Aslam2013, Weber2010, Robledo2011, Abobeih2022, Abobeih2018, Vorobyov2013}.
\begin{figure}[b]
    \centering
    \includegraphics[width=0.5\linewidth]{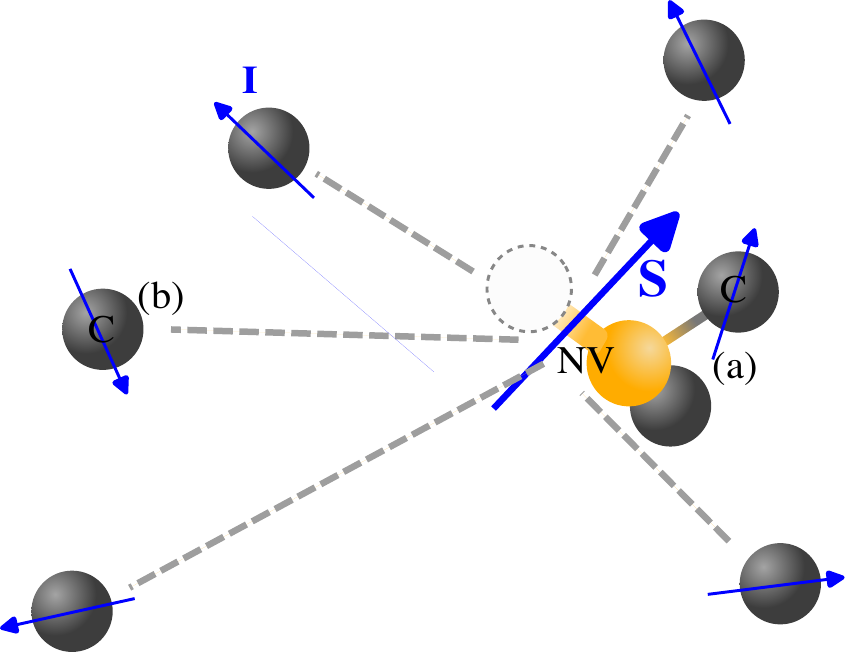}
    \caption{The nitrogen-vacancy (NV) center is a single defect in diamond. It consists of an empty site (vacancy) in the diamond lattice next to a nitrogen atom replacing one carbon atom. When negatively charged and in its ground state, the electron spin (S) forms a spin triplet while the spin of the nitrogen atom depends on the type of the nitrogen isotope and amounts to $S=1/2$ ($S=1$) in case of a $^{15}$N ($^{14}$N) atom. By hyperfine interaction, the electron spin couples to the nuclear spins ($I$) of surrounding carbon $^{13}$C atoms in the host diamond lattice, forming a quantum register that composes both strongly coupled (a) and weakly coupled (b) carbon spins.}
    \label{fig:NVcenter}
\end{figure}
A qubit-qubit or qubit-qutrit platform arises from the hyperfine interaction between the electron spin of the defect and the nuclear spin of the intrinsic nitrogen atom. When negatively charged and in its ground state, the electron spin forms a spin triplet while the spin of the nitrogen atom depends on the type of the nitrogen isotope and is $S=1/2$ ($S=1$) in case of a $^{15}$N ($^{14}$N) atom. Scalability is possible by coupling the electron spins of two NV centers, where the creation of entanglement is based on dipolar interaction, photon-, or phonon-mediated \cite{Kubo2010, Neumann2010, Auer2016, Burkard2017, Fukami2021, Hannes2024, Liu2024, Javadzade2024}. In terms of register size, the most successful approach for the realization of a larger quantum register with ultra-long coherence times, however, consists of controlling additional nuclear spins of $^{13}\rm C$ atoms in the host diamond material which also couple to the central electron spin and occur in diamond with a 1.1\% natural abundance.
Full coherent control of up to eight $^{13}\rm C$ \cite{Bradley2019} and sensing of up to 50 \cite{Abobeih2019, Stolpe2024} nuclear spins has been realized in experiments.
For implementing larger and deeper quantum circuits with the long-term goal of fault-tolerant computation, high-fidelity control of the $^{13}\rm C$ nuclear spin qubits is an essential prerequisite, which is a motivation for this work.
While the decoherence of the central electron spin due to the nuclear spin bath has been addressed in recent work \cite{Casanova2016, Maile2023, Chen2023, Bartling2023, Ungar2024, Bartling2024}, coherent errors on the nuclear spin qubits arise due to unwanted spin-flips when off-resonantly driving transitions in the dense energy spectrum of the hyperfine-coupled system. Additionally, the relative positioning of the nuclear spin axis towards the main axis of the NV center remains a challenge in state-of-the-art gate schemes, limiting the choice of $^{13}\rm C$  nuclear spins suitable as additional qubits in the register.

In this paper, we present protocols that enable the complete suppression of both off-resonant driving and off-axis hyperfine errors simultaneously. By doing so, we build on recently presented protocols for high-fidelity control of the electron spin \cite{Finsterhoelzl2024} as well as on an analysis of the fidelity created when entangling the $^{13}\rm C$ nuclear spin qubits of remote, photon-mediated registers \cite{Hannes2024}.
Our approach enables operations in the strong driving regime, leading to advantageous gate times, while simultaneously allowing for a complete suppression of both types of errors.
We show that this is possible by exploiting synchronization effects between resonantly and off-resonantly driven transitions \cite{Koch2022, Glaser2015, Sattler1999, Alexander1961, Russ2018, Heinz2021, morton2006, Noiri2022, Schirmer2005} -- which in case of the NV center occur for certain choices of intensity and frequency of the external AC and DC magnetic fields \cite{Jiang2009, Robledo2011, Zhao2012_1, Finsterhoelzl2024} -- where we focus on the state-of-the-art gate scheme for conditioned operations on the nuclear spins determined by the state of the electron spin which consists of an interplay between dynamical decoupling (DD) pulses on the electron spin and radio-frequency (rf) pulses on the nuclear spins \cite{Bradley2019}.

The remaining sections of this paper are organized as follows. In Sec.~\ref{sec:model}, we present the model we use to describe the NV system coupled to $^{13}\rm C$ nuclear spins. In Secs.~\ref{sec1}-\ref{sec3}, we review the protocol of the DDrf two-qubit entangling operation between the central electron spin and a coupled nuclear spin and explain how synchronizing effects may be included to eliminate errors due to off-resonant driving. To this end, we show in Sec.~\ref{sec2} how to use synchronization effects to circumvent unwanted spin-flips due to the off-resonant driving of unwanted transitions. We demonstrate the applicability of our method when the perpendicular hyperfine coupling component of the coupled $^{13}$C spin is negligible compared to the parallel component ($A_\perp^{13C} \ll A_\parallel^{13C}$. Next, in Sec.~\ref{sec3}, we extend the analysis to the case of a non-vanishing perpendicular hyperfine coupling between the electron spin and a $^{13}\rm C$ nuclear spin and demonstrate a protocol that can suppress both the error due to the off-resonant drive as well as due to the perpendicular hyperfine component. Finally, in Sec.~\ref{sec:conclusion}, we summarize our findings and offer perspectives for further research.

\section{Model}
\label{sec:model}
The NV center is a point defect in diamond where two carbon atoms are replaced by a $^{15}$N or $^{14}$N nitrogen atom with an empty nearest-neighbor site (see Fig.~\ref{fig:NVcenter}). In the negative charge state, the electron spin forms a spin triplet, and the hyperfine interaction couples it to the nuclear spin of the nitrogen atom with nuclear spin $I=1$ ($I=1/2$) in case of a $^{14}$N ($^{15}$N). With a natural abundance of 1.1\%, ${13}\rm C$ atoms with nuclear spin $I=1/2$ are randomly located in the host lattice. These also couple to the electron spin of the defect, while coupling between the nuclear spins is very weak against this and thus negligible. This results in the connectivity of a central-spin system where the electron spin mediates all couplings, described by the
total Hamiltonian,
\begin{equation}
 H_0=H_e + \sum_j H^j_n + \sum_j H^j_i
\label{eq:totalhamiltonian}
\end{equation}
with the electronic, nuclear, and interaction parts,
\begin{align}
H_e&=DS_z^2+\gamma_{e}B_z  S_z \label{eq:Hamiltonian1},\\
    H^j_n&=Q^j (I^j_z)^2 -\gamma^j_nB_z  I^j_z \label{eq:Hamiltonian2},\\H_i^j&=\textit{\textbf{SA}}^j\textit{\textbf{I}}^j \label{eq:Hamiltonian3},
\end{align}
where we have chosen energy units such that $\hbar \equiv 1$. Eq.~\eqref{eq:Hamiltonian1} describes the free evolution of the electron spin, Eq.~\eqref{eq:Hamiltonian2} the free evolution of the nuclear spin with the index $j$ where $j$ runs over all coupled nuclear spins, and Eq.~\eqref{eq:Hamiltonian3} the hyperfine interaction between the electron and the $j$th nuclear spin given by the hyperfine tensor $\textit{\textbf{A}}^j$. 
Here, $D/(2\pi)=2.88\,\text{GHz}$ denotes the zero-field splitting separating the electronic spin energy levels $m_s=0$ and $m_s=\pm 1$ when $B_z=0$, $Q^j$ denotes the respective quadrupole splitting, $B_z$ denotes the $z$-component of the static magnetic field $\textbf{B}=(0, 0, B_z)$ which lifts the degeneracy of the spin levels and is applied in the direction of the system's main axis, and $\textit{\textbf{S}}=(S_x,S_y,S_z)$ is the electronic spin operator with $S=1$ and $\textit{\textbf{I}}^j=(I^j_x,I^j_y,I^j_z)$ the nuclear spin operator. The reduced electronic gyromagnetic ratio is given by $\gamma_{e} = \mu_B g_e$ with $\mu_B/h=14.00\,\text{GHz/T}$ and $g_e=2.00$ while the reduced nuclear gyromagnetic ratio is given by $\gamma_{n}^j=\mu_N g^j_n$ with the nuclear magneton $\mu_N/h=7.63\,\text{MHz/T}$ and $g_n^j$ the respective nuclear g factor \cite{Everitt2014}.
In this work, we consider an NV center coupled to one carbon-13 atom, where we focus on the coherent control of the combined dynamics between the center electron spin and the remote nuclear $^{13}C$ spin, thus $j=C$ in Eqs.~\eqref{eq:Hamiltonian2}, \eqref{eq:Hamiltonian3}. The reduced gyromagnetic ratio is given by $\gamma_n^C/(2\pi)=10.705\,\text{MHz}$ \cite{Zaiser2019}, while $^{13}C$ does not exhibit a quadrupole splitting, $P^{C}=0$. With this, Eq.~\eqref{eq:Hamiltonian2} becomes $H_n^C= B_z \gamma_n^C$.
The interaction part between the nuclear and electronic spin given by $H_i^C=\textbf{\textit{SA}}^C\textbf{\textit{I}}^C$ in Eq.~\eqref{eq:Hamiltonian3} may be simplified to two non-zero components $A^C_{zz}\equiv A^C_{||}$ and $A^C_{zx} \equiv A^C_\perp$, based on the secular approximation \cite{Childress2006}. The values of these components depend on the position and relative axis of the respective $^{13}$C isotope towards the main axis of the NV center. With this, Eq.~\eqref{eq:totalhamiltonian} reduces to

\begin{equation}
       H_0 = D S_z^2 + B_z \gamma_{el} S_z 
       + B_z \gamma_n^C I_z^C + A_{||} S_z I^C_z + A^C_\perp S_z I^C_x.
       \label{eq:model}
\end{equation}

\section{DDRF gate scheme in the weak-driving limit}
\label{sec1}
Pioneering experiments have focused on the coherent manipulation of the electron spin and strongly coupled nuclear spins where $A_\parallel \geq 1/T_2^*$. Here, conditional logic is achieved by driving the desired transition between the hyperfine-split energy levels with a microwave or radio-frequency (rf) pulse \cite{Childress2006, Dutt2007, Robledo2011, Fuchs2011, Pfaff2013, Dolde2014, Waldherr2014, Rong2015}. A breakthrough in the coherent manipulation of weakly coupled carbon spins was achieved in 2012 by three groups independently \cite{Taminiau2012, Kolkowitz2012, Zhao2012_2}. 
In their work, a protocol was presented that is based on dynamical decoupling (DD) sequences on the electron spin and at the same time builds on the dependence of the axis and frequency of the Larmor precession of the individual carbon spins on the transversal components of their respective hyperfine tensors.
Conditional operations between electron and nuclear spins are achieved by an appropriate choice of the time delay $\tau$ between the pulses, thereby exploiting first, that the hyperfine interaction is only effective in the electron $m_s=\pm 1$ states and second, that the hyperfine tensor depends on the relative positioning of the nuclear spin toward the main axis of the center and thus differs for each individual nuclear spin. The downsides of the scheme are that it builds upon a non-negligible nuclear transversal hyperfine component and possibly long gate times.  

Recent research has extended the analysis and flexibility of nuclear spin sensing and control \cite{Vandersar2012, Casanova2015, Wang2016, Abobeih2018, Bradley2019, Bartling2022, Takou2022, Takou2023, Vanommen2024}.
In Ref.~\onlinecite{Bradley2019}, an experimentally demonstrated scheme combines both techniques by using dynamical-decoupling sequences on the electron spin with radio-frequency pulses on the nuclear spin (DDrf). However, this scheme relies on the transversal hyperfine coupling being negligibly small. To avoid errors due to the off-resonant driving of unwanted transitions in the crowded spectrum, these sequences are operated in the weak-driving regime with a Rabi frequency in the range of a few kHz. 

In this paper, we propose DDrf schemes that avoid errors due to off-resonant driving completely. By doing so, we contribute to recent proposals to enhance the flexibility and gate fidelities of the DDrf approach \cite{Vanommen2024}.
Additionally, we present a scheme that allows one to address strongly coupled nuclear spins in the presence of a transversal coupling component. To suppress errors due to unwanted spin flips, rather than avoiding the driving of unwanted transitions, we propose to make use of synchronization effects between resonantly and off-resonantly driven transitions.
With this, we build on a technique that has emerged in the field of nuclear magnetic resonance spectroscopy \cite{Koch2022, Glaser2015, Sattler1999, Alexander1961} and has also been shown theoretically \cite{Russ2018, Heinz2021} as well as demonstrated experimentally \cite{Noiri2022} for semiconductor spin qubits.

In what follows, we briefly introduce the DDrf scheme \cite{Bradley2019}. Starting from Eq.~\eqref{eq:model}, an rf-driving field with frequency $\omega$ is introduced that is polarized perpendicular to the main axis of the NV center along the x-axis. It has a constant driving strength $B_1$ and phase $\phi$ according to 
\begin{equation}
    H_d=2 B_1 \cos{(\omega t+\phi)} I_x.
    \label{eq:drive}
\end{equation}
The computational space of the encoded qubits is defined using the electron spin states $|m_s=0\rangle \equiv |0\rangle_e$ and $|m_s=-1\rangle \equiv |1\rangle_e$ and the nuclear spin states $|m_l=+\nicefrac{1}{2}\rangle \equiv |0\rangle_n$ and $|m_l=-\nicefrac{1}{2}\rangle \equiv |1\rangle_n$. The $\ket{m_s=+1}$ state of the electron spin triplet is assumed to be far detuned, thus $B_1\ll \delta$ with $\delta = E_{m_s=-1}-E_{m_s=+1}$ which is given by $\delta = 2\gamma_e B_z$ if $\gamma_e B_z < D$ and $\delta = 2D$ if $\gamma_e B_z \geq D$ \cite{Everitt2014}.
Here, only carbon spins with a negligible transversal hyperfine component are considered, thus $A_\perp^C \ll \gamma_n B_z-A_\parallel$.

Moving into the interaction picture defined by the unitary transformation $U_1=e^{it (DS_z^2+B_z \gamma_{el} S_z +\omega  I_z)}$ and $\Tilde{H}=U_1HU_1^\dagger -iU_1\delta_tU_1^\dagger$, and performing the rotating wave approximation, the total Hamiltonian in the electronic computational subspace $H=H_0+H_d$ reads as
\begin{equation}
\Tilde{H} = (\gamma_n B_z-\omega) I_z - A_{||}/2 (\mathrm{I}_e-\sigma^e_z) I_z 
+ B_1(\sin\phi I_x+\cos\phi I_y).
\label{eq:hrwa}
\end{equation}
A C$_e$NOT$_n$ gate is achieved by driving the nuclear spin on resonance when the electron is in the $\ket{1}$ state, at the frequency $\omega=\gamma_n B_z-A_{||}$. Equation~\eqref{eq:hrwa} may be rewritten in terms of two parts containing the action on the nuclear spin in the respective subspace of the electronic spin
\begin{equation}
    \Tilde{H} = H^n_0 \ket{0}_e\bra{0}_e + H^n_1 \ket{1}_e\bra{1}_e,
    \label{eq:parthamiltonians}
\end{equation}
with $H_1^n=B_1(\sin\phi I_x+\cos\phi I_y)$ and $H^n_0=A_{||}^C I_z$. Here, the driving field is assumed to act only in the resonantly targeted $\ket{1}_e$ electron subspace, effectively neglecting effects due to off-resonantly driving the nuclear spin in the $\ket{0}_e$ subspace.
During the duration of the controlled operation of the nuclear spin, the electronic spin is subjected to $N$ regularly spaced, nearly instantaneous $\pi$ pulses in the GHz regime. The pulse sequence is given by $(\tau - \pi - 2\tau - \pi - \tau)^{N/2}$ with $N$ even and $\tau$ the interpulse spacing. Note that the total gate time is thus given by $t_g=2N\tau$.

The interpulse delay $\tau$ is chosen to be an integer multiple of the period of the Larmor precession of the carbon nuclear spins,
\begin{equation}
\tau=l\tau_L,
\label{eq:taucondition}
\end{equation}
with $l = 1,2,3,\ldots$, $\tau_L=2\pi/\omega_L$, and $\omega_L=\gamma_n^CB_z$, leading to a decoupling of unwanted interactions between the electron and environmental nuclear spins \cite{Childress2006, Abobeih2018}. 

The time evolution operator $V$ acting on the nuclear spin may now be separated into two parts depending on the state of the electron spin $V_0$ and $V_1$ given by
\begin{equation}
    V= \ketbra{0}{0}V_0 +  \ketbra{1}{1}V_1,
    \label{eq:v}
\end{equation}
where
%
\begin{eqnarray}
V_0 &=& U_0(\tau)  U_1(2\tau,\phi_{K-1})  U_0(2\tau) \dots 
 U_0(2\tau)  U_1(2\tau,\phi_{2})  U_0(\tau),\\  
\label{eq:v0}
V_1 &=& U_1(\tau, \phi_K)  U_0(2\tau)  U_1(2\tau,\phi_{K-2}) \dots 
 U_1(2\tau,\phi_3) 
U_0(2\tau)  U_1(\tau,\phi_{1}),
\label{eq:v1}
\end{eqnarray}
%
where $k$ labels the sequences of duration $\tau$ during which the driving field Eq.~\eqref{eq:drive} of phase $\phi_k$ acts on the system, thus $k=1, \dots K$, $K=N+1$. The unitaries $U_0(t)$ and $U_1(t,B_1,\phi)$ are given by
 \begin{equation}
             U_0(t)= e^{-iH^n_0t}
         =\begin{pmatrix}
        e^{-iA_{||}t/2} & 0 \\
        0 & e^{iA_{||}t/2} \\
    \end{pmatrix},
 \end{equation}
 and
\begin{equation}
            U_1(t,B_1,\phi)= e^{-iH^n_1t} 
    = \begin{pmatrix}
        \cos(B_1t/2) & -i\sin(B_1t/2)e^{-i\phi} \\
        -i\sin(B_1t/2)e^{i\phi} & \cos(B_1t/2)
    \end{pmatrix}.
\end{equation}
Note that k is even in $V_0$ and odd in $V_1$. 
The conditioned operation is  achieved by adding $\pi$ only to the phase of the odd pulses acting on the nuclear spin when the electron is in the $\ket{1}$ state and a phase shift is introduced to all pulses determines the axis of the rotation of the nuclear spin, yielding $\phi_k=\varphi+\phi'_k$ with 
\begin{equation}
\phi'_k = 
    \begin{cases}
        (k-1)\phi_\tau +\pi, & \text{if $k$ odd},\\
        (k-1)\phi_\tau, & \text{if $k$ even}.
    \end{cases}
    \label{eq:phik}
\end{equation}
Substituting Eq.~\eqref{eq:phik} into Eqs.~\eqref{eq:v0}--\eqref{eq:v1} and defining the rotation operators acting on the nuclear spin subspace with $R_z(\phi)=\exp{\left[-i\theta I_z\right]}$ and $R_\varphi(\phi)=\exp{\left[-i\theta (\cos{\varphi}I_x+\sin{\varphi}I_y) \right]}$ results in 
\begin{eqnarray}
    V_0&=R_z(NA_{||}^C\tau)  R_\phi(NB_1\tau), \label{eq:v0n}\\
    V_1&=R_z(NA_{||}^C\tau)  R_\phi(-NB_1\tau).
    \label{eq:v1n}
\end{eqnarray}
Combining Eqs.~\eqref{eq:v0n} and \eqref{eq:v1n} with Eq.~\eqref{eq:v} results in a total time evolution operator containing an unconditional rotation of the nuclear spin around the $z$-axis, $V_z= \mathrm{I}_e \otimes R_z(NA_{||}^C\tau)$, and a conditional rotation of the nuclear spin $V_{\rm crot}=\ketbra{0}{0} \otimes R_\phi(NB_1\tau)+ \ketbra{1}{1} \otimes R_\phi(-NB_1\tau)$ where the angle is given by $NB_1\tau$ and the axis by $\phi$,
\begin{equation}
    V=V_z  V_{\rm crot}.
    \label{eq:crot}
\end{equation}
Equation~\eqref{eq:crot} is equivalent to a 
C$_e$NOT$_n=\ketbra{0}{0}_e \otimes \mathrm{I}_n + \ketbra{1}{1} \otimes 2I_x$ up to single-qubit operations on the electron as well as the nuclear spin by choosing 
\begin{equation}
    NB_1\tau=\pi/2,
    \label{eq:condition0}
\end{equation}
and $\varphi=0$, resulting in
\begin{equation}
    V_{\rm crot}=e^{i\pi/4} \left(R^e_z(\pi/2) \otimes \mathrm{I}_n\right) \left( \mathrm{I}_e  R^n_x(\pi/2) \right)  {\rm C}_e{\rm NOT}_n.
    \label{eq:vcrot2}
\end{equation}
Typical driving strengths are in the $\sim$ kHz regime, resulting in gate times of $\sim 1 \,{\rm ms}$.

\section{Minimizing unwanted spin-flips in fast C$_e$NOT$_n$ gates}
\label{sec2}
This section explains our DDrf-protocol, which allows high-fidelity operations beyond the weak driving limit.
Contrary to the derivation in Ref.~\onlinecite{Bradley2019}, we include the off-resonant driving of the nuclear spin in the election $\ket{0}_e$ subspace, thus Eq.~\eqref{eq:parthamiltonians} now contains the driving term in both subspaces, and therefore, $\Tilde{H} = H^n_0 \ketbra{0}{0} + H^n_1 \ketbra{1}{1}$,
with $H^n_0=A_{||} I_z+B_1(\sin\phi I_x+\cos\phi I_y)$ and $H_1^n=B_1(\sin\phi I_x+\cos\phi I_y)$. The time evolution operator of the $\ket{0}_e$ subspace now also contains a drive with the Rabi frequency given by $\Omega=\sqrt{(A_\parallel/2)^2+(B_1/2)^2}$, thus
 \begin{equation}
    U_0(t,\Omega,\phi)= e^{-iH^n_0t}=
    \begin{pmatrix}
       \cos(\Omega t/2)-i\frac{A_\parallel}{2\Omega}\sin(\Omega t) &  -i\frac{B_1}{2\Omega}\sin(\Omega t/2)e^{-i\phi} \\
         -i\frac{B_1}{2\Omega}\sin(\Omega t/2)e^{i\phi} & \cos(\Omega t/2) +i\frac{A_\parallel}{2\Omega}\sin(\Omega t)
    \end{pmatrix},
    \label{eq:u0new}
\end{equation}
and
    \begin{equation}
    U_1(t,B_1,\phi)= e^{-iH^n_1t}=
    \begin{pmatrix}
        \cos(B_1t/2) & -i\sin(B_1t/2)e^{-i\phi}  \\
        -i\sin(B_1t/2)e^{i\phi}  & \cos(B_1t/2)
    \end{pmatrix}.
    \label{eq:u1new}
\end{equation} 
With this, both $V_0$ and $V_1$ are phase-dependent and contain pulses of even and odd numbers, given by 
%
 $   V= \ketbra{0}{0}V_0 +  \ketbra{1}{1}V_1$,
%
and 
\begin{eqnarray}
V_0 &=& U_0(\tau, \phi_K)  U_1(2\tau,\phi_{K-1})  U_0(2\tau,\phi_{K-2}) \dots
 U_0(2\tau,\phi_3)  U_1(2\tau,\phi_{2})  U_0(\tau,\phi_1)  ,
\label{eq:v0new}\\
V_1 &=& U_1(\tau, \phi_K)  U_0(2\tau,\phi_{K-1})  U_1(2\tau,\phi_{K-2}) \dots
 U_1(2\tau,\phi_3)  U_0(2\tau,\phi_2)  U_1(\tau,\phi_{1}),
\label{eq:v1new}
\end{eqnarray}
with $\tau=l\tau_L$ as in Eq.~\eqref{eq:taucondition} and explained in Sec.~\ref{sec1}.

\begin{figure}
    \centering
    \includegraphics[width=0.6\linewidth]{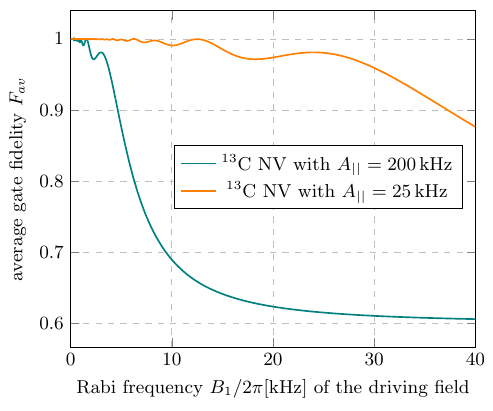}
    \caption{Average gate fidelity $F_{\rm av}$ (Eq.~\eqref{eq:gatefidelity}) of a C$_e$NOT$_n$-gate between the electron spin of an NV center and a $^{13}C$ nuclear spin as a function of the Rabi frequency of the driving field $B_1$ for exemplary values of the parallel coupling component $A_{||}$ between the nuclear and electronic spin.}
    \label{fig:fid1}
\end{figure}

We evaluate the error emerging due to the off-resonant drive by calculating the average gate fidelity defined by 
\begin{equation}
    F_{\rm av}(U_{\rm CNOT},V_{\rm CNOT})=\frac{d+\left| \textup{Tr}\left(U_{\rm CNOT}^\dagger,V_{\rm CNOT}\right) \right|^2}{d(d+1)},
    \label{eq:gatefidelity}
\end{equation}
with $d=4$ the dimension of the Hilbert subspace and with the actual C$_e$NOT$_n$ given by
\begin{equation}
    V_{\rm CNOT}=
    e^{-i\pi/4} \left(R^e_z(-\pi/2) \otimes \mathrm{I}_n\right)  \left( \mathrm{I}_e \otimes R^n_x(-\pi/2) \right)  V^\dagger  V_{\rm CROT},
    \label{eq:uact}
\end{equation}
cf.\ Eq.~\eqref{eq:crot} and Eq.~\eqref{eq:vcrot2}, and the ideal C$_e$NOT$_n$ is given by
$U_{\rm CNOT}=V= \ketbra{0}{0}\openone +  \ketbra{1}{1}\sigma_x$.
%
%
 In Fig.~\ref{fig:fid1}, we plot the resulting average gate fidelity $F_{\rm av}$ for exemplary values of the hyperfine coupling $A_\parallel$ as a function of the Rabi frequency $B_1$. The fidelity strongly depends on the driving strength $B_1$. In the weak driving regime where $B_1 \ll A_\perp$, high-fidelity operations with $F_{\rm av} \approx 1$ are possible. However, with increasing driving strength, the fidelity deteriorates and deviates significantly from unity. This is due to unwanted spin flips from the off-resonantly driven $\ket{00}\rightarrow \ket{01}$ transition where the nuclear spin is flipped despite the electron spin being in $\ket{0}$. 
In the following, we present a protocol that suppresses this error.
The key idea of the scheme is that the error due to the off-resonant drive is canceled out by choosing the driving strength $B_1$ such that a synchronization of the Rabi frequencies of both the resonantly and the off-resonantly driven transitions is achieved: With this, the Bloch vector of the off-resonant transition has to perform a $2\pi$ rotation around an axis in the X-Z plane during the gate time $t_g$. In the case of the NV center, this axis is determined by $B_1$ and $A_\parallel$ \cite{Finsterhoelzl2024}. 

\begin{figure}
    \centering
    \includegraphics[width=\linewidth]{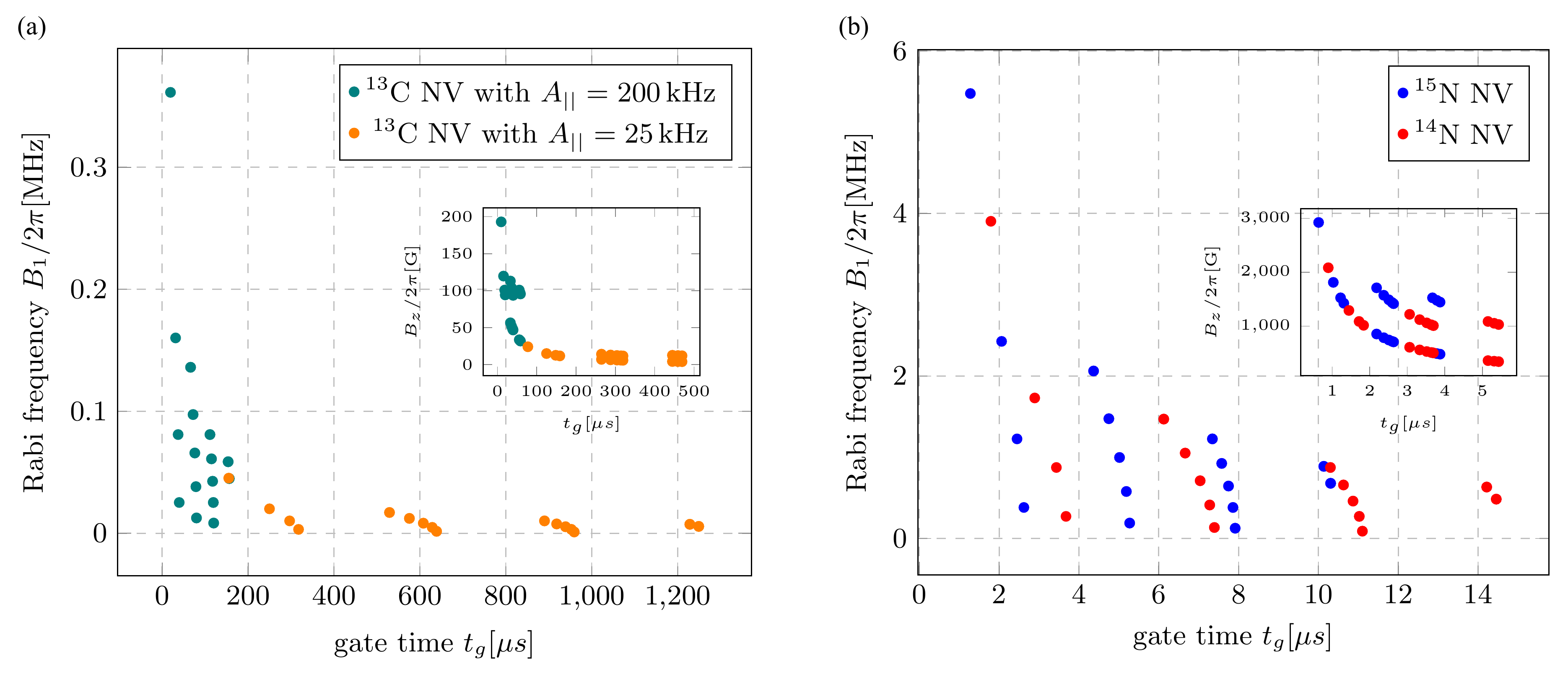}
    \caption{Driving strength $B_1/2\pi$ as a function of the gate time $t_g$ for several choices of $B_1^{\rm sync}$ for the electron spin of an NV-center coupled to the nuclear spin of a $^{13}C$ atom with different exemplary coupling strengths $A_{||}$ (a), or to the nuclear spin of the intrinsic $^{14}N$/$^{15}N$ atom (b). The inset plot depicts the respective minimal value for the static external magnetic field $B_z^{\rm sync}/2\pi$ (units of Gauss) corresponding to one set of parameterswhere $B_z^{\rm sync}=B_z^{\rm sync}(B_1^{\rm sync},n,N,\tau)$, and the related gate time $t_g$. Higher values for the external field may be achieved by taking integer multiples of these minimal values according to Eq.~\eqref{eq:bz}.}
    \label{fig:gatetimeNV}
\end{figure}

This scheme may be incorporated into the DDrf gate by synchronizing the action of the unitaries $U_0(\tau)$ (Eq.\eqref{eq:u0new}) and $U_1(\tau)$ (Eq.\eqref{eq:u1new}) on the coupled system during each pulse interval of length $\tau$. Here, $U_0(\tau)$, $U_1(\tau)$ describe the time evolution on the nuclear spin subspace $\ket{0}$, $\ket{1}$ respectively, cf. Eqs.~\eqref{eq:v0new} and \eqref{eq:v1new}.
To this end, values for $B_1$ are chosen such that with this condition, both Bloch vectors are synchronized in the above-described way at the end of each interpulse spacing $\tau$. This is the case for all values for $B_1$ given by
\begin{equation}
    \Omega \tau = \sqrt{(A_\parallel/2)^2+(B_1/2)^2}\tau = m \pi,
    \label{eq:condition1}
\end{equation}
with $m =1,2,3,\ldots$. If $m$ is odd, then $N/2$ has to be an even integer to avoid a relative phase shift. This results in 
\begin{eqnarray}
    V_0=&R_\varphi(NB_1\tau), \label{eq:v0nn}\\
    V_1=&R_\varphi(-NB_1\tau),
    \label{eq:v1nn}
\end{eqnarray}
leading to a total time evolution according to  
\begin{eqnarray}
    V_{\rm sync}=V_{\rm CROT},
    \label{eq:vsync}
\end{eqnarray}
where $V_{\rm CROT}$ is given in Eq.~\eqref{eq:vcrot2}. Note that when choosing $V_z=\openone$, Eq.~\eqref{eq:vsync} is equivalent to Eq.~\eqref{eq:crot} in the scheme described in Sec.~\ref{sec1}, since the synchronization condition takes care of the correction of single-qubit rotations of the nuclear spin. 
Setting
\begin{equation}
    NB_1\tau=(2n+1)\pi/2,
    \label{eq:condition2}
\end{equation}
with $n =0,1,2,3,\ldots$ results in a CNOT gate described by Eq.~\eqref{eq:uact}.
Combining Eqs.~\eqref{eq:condition1} and \eqref{eq:condition2} results in the condition for the driving field as a function of the parallel hyperfine component $A_{||}^C$,
\begin{equation}
    B_1^{\rm sync}(n,m)=A_{||}^C\sqrt{\frac{(2n+1)^2}{4m^2N^2-(2n+1)^2}},\quad m,n =1,2,3,/ldots.
    \label{eq:sync}
\end{equation}
Note that with Eq.~\eqref{eq:condition2}, this synchronization condition also fixes the values for the interpulse spacing $\tau$.
To fulfill Eq.~\eqref{eq:taucondition} which determines $\tau$ as an integer multiple of the $^{13}$C nuclear spin Larmor precession period (cf.\ Sec.~\ref{sec1}), the external static magnetic field $B_z$ has to be adapted accordingly. Appropriate values are determined by 
\begin{equation}
    B_z^{sync}(B_1^{sync},n,N,\tau) = \frac{l}{\tau \gamma_n} = \frac{l}{2n+1} \frac{4NB_1^{sync}}{\gamma_n}.
    \label{eq:bz}
\end{equation}
The total gate time for the synchronized scheme is given by
\begin{equation}
    t_g (A_\parallel,N,m,n) = \frac{\pi}{A_\parallel }\sqrt{N^2m^2 - \frac{(2n+1)^2}{4}}.
    \label{eq:syncgatetime}
\end{equation}
The fastest gate is achieved with $n=7$, $m=1$, and $N=2$, thus with a rotation of the nuclear spin by an angle of $7\pi$. This holds independently of the coupling strength $A_\parallel$ and contrasts synchronization schemes targeting the electron spin \cite{Finsterhoelzl2024}, where the fastest gate is achieved with a $\pi$-rotation of the driven transition ($n=0$).
For strongly (weakly) coupled carbon spins, $A_\parallel/2\pi \sim 200\,\rm kHz$ ($A_\parallel/2\pi \sim 25\,\rm kHz$), this results in a Rabi drive $B_1/2\pi=361\,\rm kHz$ ($B_1/2\pi\approx45.2\,\rm kHz$) and in the respective gate time $t_g\approx19.3\,\mu\rm s$ ($t_g\approx 155\,\mu\rm s$).
Other selected values of the driving field and the respective gate time for different choices of $n, m, N$ are plotted in Fig.~\ref{fig:gatetimeNV}(a). 
The inset plot depicts the respective minimal value for the static external magnetic field $B_z/2\pi$, corresponding to one set of parameters and the related gate time $t_g$. Stronger regimes of the field strength may be achieved by taking integer multiples of these minimal values according to Eq.~\eqref{eq:bz}. Values in recent experiments operate in regimes between $\sim50\,\rm G$ to $\sim2000\,\rm G$ \cite{Bradley2019,Bradley2021}.

Figure~\ref{fig:gatetimeNV}(b) gives the related numbers for an NV-center composed of an $^{15}N$ ($^{14}N$) nitrogen atom. We include this in the presentation as our protocol is easily adaptable to this spin type intrinsic to the NV center \cite{Finsterhoelzl2024}. Here, electron-nuclear gates with a gate time of $t_g\approx0.896\,\mu\text{s}$ ($t_g\approx0.639\,\mu\text{s}$) are possible for the $^{14}N$ ($^{15}N$) NV-center. In this case, the static magnetic field $B_z$ is in the strong regime with $\geq 2000\,\rm G$. This constitutes a significant speedup compared to gate times of $\sim 0.4-2.4\,\text{ms}$ with a driving field strength of $\sim 1\,\text{kHz}$ in recent experiments \cite{Bradley2019}.

\subsection{Choices for the static magnetic field component $B_z$}
While the external magnetic field strength depends on the chosen synchronization sequence, the possibility for a constant $B_z$ while addressing different nuclear spins sequentially exists. If two different coupled nuclear spins with hyperfine components $A_{\parallel,1}, A_{\parallel,2}$ are addressed successively, the external magnetic field $B_z$ may be kept constant in case the ratio $x\equiv A_{\parallel,1}/A_{\parallel,2}$ fulfills the condition
\begin{equation}
    x = \frac{l_2}{l_1} \left[\frac{m_1^2-\frac{(2n_1+1)^2}{4N_1^2}}{m_2^2-\frac{(2n_2+1)^2}{4N_2^2}}\right]^{1/2},
    \label{eq:bzconstant}
\end{equation}
where $N_i,n_i,m_i$ are given for nuclear spin $i=\{1,2\}$ by Eq.\eqref{eq:sync} and $l_i$ by Eq.\eqref{eq:bz}.
With the allowed combinations of $N_i,n_i,m_i,l_i$, $i\in \{1,2\}$, the ratio of two targeted spins may be approximated by $x$ with a precision depending on the chosen values. For instance, by allowing integer numbers in $\{1\dots 10\}$ for all integer values in Eq.~\eqref{eq:bzconstant}, and neglecting double counting, roughly $10^8$ values between $x_{\rm min}\approx 0.00347$ and $x_{\rm max}\approx86.215$ are representable. At the same time, many of them are tunable with a precision of $10^{-5}$.

\subsection{C$_e$NOT$_n$ in the presence of a perpendicular hyperfine tensor component}
\label{sec3}

Here, we present a protocol that predicts a vanishing gate error in the presence of a non-negligible perpendicular hyperfine tensor component $A_\perp$. Our approach relies on detuning the frequency of the driving laser field relative to the energy of the targeted transition. This leads to a dependency of the driving strength on the phase of the laser pulse, resulting in different values for $\tau_i$ and thus for the static field $B_z$ if included in the DDrf scheme. For this reason, we present the protocol for a C$_e$NOT$_n$ operation without DD sequences suitable for strongly coupled nuclear spins.

Starting from the Hamiltonian in Eq.~\eqref{eq:totalhamiltonian} and the driving field in Eq.~\eqref{eq:drive}, we now include $A_\perp$ in the calculation and transform Eq.~\eqref{eq:totalhamiltonian} with the unitary $U_2=e^{i\omega t (\sin{\beta} I_x+\cos{\beta} I_z)}$ and $\Tilde{H}=U_2HU_2^\dagger -iU_2\delta_tU_2^\dagger$, with $\omega_1=\sqrt{A_\perp^2+(\omega_l-A_\parallel)^2}$ the frequency of the nuclear spin precession when the electron spin is in the $m_s=-1$ state with a tilted axis given by $\cos{\beta} I_z+\sin{\beta} I_x$ and $\sin{\beta}=\frac{A_\perp}{\omega_1}$, $\cos{\beta}=\frac{\omega_l-A_\parallel}{\omega_1}$. 
Dropping fast oscillating terms, the nuclear Hamiltonian in the respective electronic subspace in Eq.\eqref{eq:parthamiltonians} becomes 
\begin{multline}
    H^n_0 = \left[ (\omega_l \cos{\beta}-\omega)\cos{\beta} - \Omega \sin{\beta} \cos{\beta} \cos{\phi} \right]I_z\\
    +\left[ (\omega_l \cos{\beta}-\omega)\sin{\beta} + \Omega \cos{\beta}^2 \cos{\phi} \right]I_x
    +\left[ \Omega \cos{\beta} \sin{\phi} \right]I_y,
    \label{eq:Haperp0}
\end{multline}
and
\begin{multline}
    H^n_1  =\left[ (\omega_1 \cos{\beta}-\omega)\cos{\beta} - \Omega \sin{\beta} \cos{\beta} \cos{\phi} \right]I_z  \\ 
    +\left[ (\omega_1 \cos{\beta}-\omega)\sin{\beta} + \Omega \cos{\beta}^2 \cos{\phi} \right]I_x
    + \left[ \Omega \cos{\beta} \sin{\phi} \right]I_y.
    \label{eq:Haperp1}
\end{multline}
We choose the detuning to fulfill 
\begin{equation}
    \omega_1-\omega = \Omega \sin{\beta}\cos{\phi}.
\end{equation}
A C$_n$NOT$_e$ is applied by setting $\phi=0$ and demanding that $t_g=\frac{(2n+1) \pi}{\Omega}$, $n =0,1,2,\ldots$, cf.\ Sec.~\ref{sec2}. To achieve synchronization, the condition $\Omega_0 tg=m\pi$ with $m=0,1,2,\ldots$ and $U_0^2=\left[ (\omega_l \cos{\beta}-\omega)\cos{\beta} - \Omega \sin{\beta} \cos{\beta} \cos{\phi} \right]^2+\left[ (\omega_l \cos{\beta}-\omega)\sin{\beta} + \Omega \cos{\beta}^2 \cos{\phi} \right]^2+\left[ \Omega \cos{\beta} \sin{\phi} \right]^2$ has to be fulfilled additionally (see Sec.~\ref{sec2}) which determines the driving strength $\Omega$. 
Substituting these conditions into Eq.~\eqref{eq:Haperp0}-\eqref{eq:Haperp1}, the time evolution operator of the entire sequence can be expressed as
$U= \ketbra{0}{0}U_0 +  \ketbra{1}{1}U_1$
with
 \begin{equation}
    U_0 = e^{-iH^n_0t_g}= (-1)^m \mathbb{1}, \quad U_1= e^{-iH^n_1t_g}=-i\sigma_x,
\end{equation} 
which is locally equivalent to C$_n$NOT$_e$.

\subsection{Beneficial parameter regimes}
Lastly, we discuss the parameter regimes for which our protocols predict an enhancement of the gate fidelity even when considering the errors due to the RWA. 
In recent experiments, the main coherent errors limiting the gate fidelity of $^{13}\rm C$-NV entangling operations have been identified as the off-resonant drive of neighboring transitions and a non-negligible hyperfine tensor perpendicular component $A_\perp$ \cite{Bradley2019}. We check the validity of the rotating wave approximation by simulating a synchronized C$_e$NOT$_n$ operation for different choices of $B_{\rm sync}$ (cf.\ Eq.\eqref{eq:sync}) with the equations of motion of the full system. We also use the simulation results to evaluate the impact of the above-mentioned errors in different operation regimes. Tables~\ref{tab:regimevalues} and \ref{tab:aperpvalues} list the resulting gate fidelities for strong and weak driving, various choices of the static magnetic field, and exemplary values of the parallel and perpendicular hyperfine tensor components $A_\parallel$, $A_\perp$.
Here, 
gate fidelities are enhanced despite errors due to the approximation in the regime where the RWA is valid, $B_1^{\rm sync}\ll \|\omega_l-A_\parallel\|$, and in case the perpendicular hyperfine component $A_\perp$ is of a similar order of magnitude as $A_\parallel$. In this case, the synchronization protocol leads to a strong enhancement of the gate fidelity, even when taking the errors due to the approximation into account.
However, in the regime where the energy of the driven transition is in the order of magnitude of the driving strength, $|B_1^{\rm sync}|\gtrsim \|\omega_l-A_\parallel\|$, the rotating wave approximation breaks down which is apparent in low fidelities between the full and the approximated solution. 
Also, when the perpendicular component is negligible compared to the targeted transition frequency, $A_\perp \ll \|\omega_l-A_\parallel\|$, the error induced by the RWA has a similar impact on the gate performance for both protocols presented above. 
 \begin{table*}[t]
   \label{tab:regimevalues}
    \caption{Validity of the rotating wave approximation (RWA): Full simulation of the equations of motion $F_{\rm ave}$ of an entangling C$_e$NOT$_n$ operation with a $^{13}\rm C$ spin, with $n=0$, $m=1$ in Eq.~\eqref{eq:sync} -- thus, for the fastest synchronized gate with the strongest drive given by $B_1^{\rm sync}=A_\parallel/\sqrt{3}$ (without DD pulses on the electron spin) -- and different exemplary values of $A_\parallel$, $A_\perp$ and $B_z$. CNOT denotes the ideal operation. In the third column, $U_{\rm full}^{\rm sync,1}$ denotes the full simulation of the synchronized gate, and deviations of the fidelity from unity account for errors due to the neglect of terms of higher order in $\omega$ in the RWA. Clearly, in the regime where $B_1^{\rm sync}\ll \|\omega_l-A_\parallel\|$, the error remains small, confirming the validity of our protocol for this regime. In the fourth column, we add an exemplary non-negligible $A_\perp$ to the calculation, which is not accounted for in the scheme and induces a further deterioration of the fidelity. In the fifth column, $U_{\rm full}^{\rm sync,2}$ denotes the full simulation of the protocol explained in Sec.~\ref{sec3} which compensates for effects of $A_\perp$ with the same parameters giving the error due to the RWA. Thus, valid regimes for this scheme are indicated if $F_{\rm ave}(\rm CNOT,U_{\rm full}^{\rm sync,2})>F_{\rm ave}(\rm CNOT,U_{\rm full}^{\rm sync,1})$. In the rightmost two columns, $t_{g}^{\rm sync,1}$,$t_{g}^{\rm sync,2}$ denote the respective gate times.}
    \begin{ruledtabular}
    \begin{tabular}{|l|l|l|l|l|l|l|}
    \hline
        $A_\parallel^{\rm C}$ [kHz] & $\omega_l$ [kHz] &  \makecell{$F_{\rm ave}(\rm CNOT,U_{\rm full}^{\rm sync,1})$ \\ ($A_\perp^{\rm C}\approx 0$) } &  \makecell{$F_{\rm ave}(\rm CNOT,U_{\rm full}^{\rm sync,1})$ \\ ($A_\perp^{\rm C}\neq 0$) }  &  \makecell{$F_{\rm ave}(\rm CNOT,U_{\rm full}^{\rm sync,2})$ \\ ($A_\perp^{\rm C}\neq 0$) }  & \makecell{$t_{g}^{\rm sync,1}$ \\$[\mu \rm s]$} & \makecell{$t_{g}^{\rm sync,2}$ \\$[\mu \rm s]$} \\ \hline
        25 & 430 & 0.99992  & 0.99886 ($A_\perp=10$) & 0.99965 ($A_\perp=10$)& 342 & 2990 \\
        25 & 1980 & 0.99999 & 0.99995 ($A_\perp=10$) & 0.99999 ($A_\perp=10$)& 341 & 2990 \\
        25 & 27 & 0.28968 & 0.42398 ($A_\perp=10$) & 0.25727 ($A_\perp=10$)& 342 & 2990 \\ 
        ~ & ~ & ~ & ~ & ~ & ~ & ~ \\ 
        200 & 430 & 0.95159 & 0.85510 ($A_\perp=20$) & 0.74629 ($A_\perp=20$) & 43 & 41 \\ 
        200 & 1980 & 0.99931 & 0.99923 ($A_\perp=20$)& 0.99955 ($A_\perp=20$)& 43& 43 \\ 
        200 & 27 & 0.95649 & 0.89905 ($A_\perp=20$)& 0.92003 ($A_\perp=20$)& 43 & 45 \\
        ~ & ~ & ~ & ~ & ~ & ~ & ~ \\ 
        400 & 430 & 0.28537 & 0.32537 ($A_\perp=20$) & 0.28958 ($A_\perp=20$) & 21 & 21 \\ 
        400 & 1980 & 0.99679 & 0.99666 ($A_\perp=20$) & 0.99608 ($A_\perp=20$)& 21 & 21 \\ 
        400 & 27 & 0.96967 & 0.95435 ($A_\perp=20$) & 0.95929 ($A_\perp=20$)& 21 & 22 \\ 
    \end{tabular}
     \end{ruledtabular}
\end{table*}
\begin{table*}[t]
\label{tab:aperpvalues}
    \caption{Beneficial regimes: Exemplary combinations of $A_\parallel$, $A_\perp$ and $B_z$ where the scheme presented in Sec.~\ref{sec3} proves beneficial even when taking errors due to the RWA into account. All calculations are analogous to those listed in Table~\ref{tab:regimevalues}, and valid regimes are indicated if $F_{\rm ave}(\rm CNOT,U_{\rm full}^{\rm sync,2})>F_{\rm ave}(\rm CNOT,U_{\rm full}^{\rm sync,1})$. Clearly, our protocol enhances the overall fidelity significantly if $B_1^{\rm sync}\ll \|\omega_l-A_\parallel\|$ and $A_\perp$ is of the same order of magnitude as $A_\parallel$.}
    \begin{ruledtabular}
    \begin{tabular}{|l|l|l|l|l|l|l|}
        $A_\parallel^{\rm C}$ [kHz] & $\omega_l$ [kHz] &  \makecell{$F_{\rm ave}(\rm CNOT,U_{\rm full}^{\rm sync,1})$ \\ ($A_\perp^{\rm C}\approx 0$) } &  \makecell{$F_{\rm ave}(\rm CNOT,U_{\rm full}^{\rm sync,1})$ \\ ($A_\perp^{\rm C}\neq 0$) }  &  \makecell{$F_{\rm ave}(\rm CNOT,U_{\rm full}^{\rm sync,2})$ \\ ($A_\perp^{\rm C}\neq 0$) }  & \makecell{$t_{g}^{\rm sync,1}$ \\$[\mu \rm s]$} & \makecell{$t_{g}^{\rm sync,2}$ \\$[\mu \rm s]$} \\ \hline
        5 & 27 & 0.99618 & 0.78181 ($A_\perp=4$) & 0.92346 ($A_\perp=4$) & 1709 & 1834 \\
        60 & 430 & 0.99826 & 0.96908 ($A_\perp=30$) & 0.98910 ($A_\perp=30$)& 142 & 142 \\ 
        100 & 1980 & 0.99982 & 0.99748 ($A_\perp=50$) & 0.99938 ($A_\perp=50$)& 85 & 85 \\ 
        100 & 1980 & 0.99982 & 0.97711 ($A_\perp=100$) & 0.99718 ($A_\perp=100$)& 85 & 88 \\ 
        400 & 27 & 0.96967 & 0.74501 ($A_\perp=200$) & 0.85868 ($A_\perp=200$) & 21 & 25 \\
    \end{tabular}
    \end{ruledtabular}
\end{table*}

\section{Conclusions}
\label{sec:conclusion}
In this work, we have presented protocols that predict the suppression of two main limiting coherent errors -- off-resonant driving of neighboring transitions in the dense energy spectrum and non-negligible perpendicular hyperfine-tensor components -- during entangling gates between nuclear and electronic spin qubits in an NV center in diamond. We have shown that this is possible by a suitable choice of the driving strength, the static magnetic field, the gate time, and the detuning of the driving field, where the synchronization between resonant and off-resonant Rabi oscillations allows for the suppression of the effects of the off-resonant effects. Our results contribute to the goal of scalable quantum computation with color centers with error rates that enable fault-tolerant algorithm design.

\begin{acknowledgments}
We acknowledge funding from the state of Baden-W\"urttemberg through the Kompetenzzentrum Quantum Computing, projects QC4BW and KQCBW24, and from the German Federal Ministry of Education and Research (BMBF) under the Grant Agreement No.~13N16212 (SPINNING).
\end{acknowledgments}

\bibliography{bibliography.bib}

\end{document}